\documentclass[ACS,STIX1COL]{WileyNJD-v2}
\usepackage{floatrow}
\usepackage{url}

\articletype{Contributions to Plasma Physics\\17th International Workshop on Plasma Edge Theory in Fusion Devices}%

\received{XXXX}
\revised{XXXX}
\accepted{XXXX}

\begin{document}

\title{Treatment of Advanced Divertor Configurations in the Flux-Coordinate Independent turbulence code GRILLIX}

\author{Thomas Body*}
\author{Andreas Stegmeir}
\author{Wladimir Zholobenko}
\author{David Coster}
\author{Frank Jenko}

\authormark{THOMAS BODY \textsc{et al.}}

\address{Tokamak Theory, Max-Planck-Institut f\"{u}r Plasmaphysik, Garching, D-85748, Germany}

\corres{*Corresponding author\quad E-mail:~\textsf{thomas.body@ipp.mpg.de}}

\abstract{Advanced divertor configurations modify the magnetic geometry of the divertor to achieve a combination of strong magnetic flux expansion, increased connection length and higher divertor volume -- to improve detachment stability, neutral/impurity confinement and heat-channel broadening. In this paper, we discuss the modification of the Flux-Coordinate Independent (FCI) turbulence code GRILLIX \cite{stegmeir_global_2019,stegmeir_grillix:_2018} to treat generalised magnetic geometry, to allow for the investigation of the effect of magnetic geometry on turbulent structures in the edge and SOL. The development of grids and parallel operators from numerically-defined magnetic equilibria is discussed, as is the application of boundary conditions via penalisation \cite{schneider_immersed_2015}, with the finite-width method generalised to treat complex non-conformal boundaries. Initial testing of hyperbolic (advection) and parabolic (diffusion) test cases is presented for the Snowflake scenario.}


\jnlcitation{\cname{%
\author{T. Body}, 
\author{A. Stegmeir}, 
\author{W. Zholobenko}, 
\author{D. Coster}, and 
\author{F. Jenko}} (\cyear{2019}), 
\ctitle{Treatment of Advanced Divertor Configurations in the Flux-Coordinate Independent turbulence code GRILLIX}, \cjournal{Contributions to Plasma Physics}, \cvol{XXXX}}

\maketitle

\section{Introduction}

Limiting the heat- and particle-exhaust to prevent damage to the divertor \& first wall is identified as one of the principle challenges in the development of fusion power reactors \cite{turnyanskiy_european_2015}. For ITER high-power operations, the `ITER baseline divertor solution' (a single-null `\textit{conventional divertor}' (\textbf{CD}) operated at partially detached conditions \cite{kallenbach_impurity_2013, turnyanskiy_european_2015}) will require operation close to the material limits for the divertor targets \cite{pitts_physics_2011}. Operating conditions for DEMO and future fusion power reactors will be significantly outside the parameter space tested in existing experimental devices and it is uncertain whether the ITER baseline solution will be sufficient for these devices. As a contingency,  several `advanced divertor configurations' (ADCs) are currently the focus of WP-DTT1-ADC project -- including the `\textit{Snowflake divertor}' (\textbf{SF}, \cite{ryutov_geometrical_2007,kotschenreuther_magnetic_2013}), the `\textit{X-divertor}' (\textbf{XD}, \cite{kotschenreuther_magnetic_2013}), the `\textit{Super-X divertor}' (\textbf{SXD}, \cite{valanju_super-x_2009}) and a `\textit{Double-null}' (\textbf{DN}) variant of a conventional divertor. These configurations are designed to have (compared to the ITER baseline scenario) \cite{lunt_proposal_2017,turnyanskiy_european_2015} \textit{higher poloidal flux expansion} -- which increases the width of the directed heat-flux channel, to increase the area over which the heat is deposited in the divertor targets --, \textit{increased divertor parallel connection length} -- which gives more field-line-parallel distance over which heat flux can be radiatively dissipated --, and \textit{improved neutral baffling} -- which allows for higher neutral and seeded-impurity densities near the divertor targets, increasing SOL/divertor radiation while keeping core radiation at acceptable levels.\\

\section{Turbulence modelling with GRILLIX}

GRILLIX is a drift-reduced Braginskii 3D fluid turbulence code, which is particularly notable for its use of the flux-coordinate independent approach (\textbf{FCI}). For a complete description of the code, see Stegmeir et al., 2019 \cite{stegmeir_global_2019} regarding the use of FCI (including the support operator method), extension to a global\footnote{i.e. non-perturbative for all fields except for an assumption of a static background magnetic field} electromagnetic model, and the penalisation method, and Zholobenko et al., 2019 \cite{Zholobenko2019}, for extensions including ion-thermal effects and implicit heat conductivity. This paper focuses on modifications to the grid generation method and the establishment of parallel operators for numerical equilibria, and a modification which allows for the use of a minimal width penalisation $\chi$ function. Although these developments are made in the context of the GRILLIX code, they are generally applicable to any turbulence code employing FCI.

The complex magnetic geometry of advanced divertor configurations can introduce challenges for codes employing field- or flux-aligned coordinates. Firstly, each of the (or, high order) X-points require extraordinary treatment, to avoid the introduction of coordinate singularities. Secondly, since turbulence occurs on roughly isotropic scales, resolving turbulent dynamics in regions of high flux-expansion (i.e. near poloidal-field nulls, or near flux-expanded strike points) requires very high resolution around the midplane. This can introduce a severe CFL criterion, which requires a greatly reduced timestep (particularly for explicit numerical schemes).

In contrast, the flux-coordinate independent method uses a grid which is independent of the magnetic field and instead encodes the magnetic field structure in the parallel operators. This approach has both advantages and disadvantages compared to the more typical field- or flux-aligned coordinates. Coordinate singularities are avoided, and X-points, O-points and the separatrix can be treated without changing the numerical method. The grid resolution may be set freely, allowing for good resolution in regions of high-flux expansion and with a minimally-restrictive CFL criterion. Additionally, the method allows for the treatment of 3D equilibria such as RMPs and stellarators \cite{shanahan_fluid_2019} -- although currently GRILLIX assumes an axisymmetric field.

Conversely, the interpolation routines incur a (slight) computational cost outright, but more significantly this also means that efficient parallelisation is more complex since the concept of up and downwind parallel `neighbours' is not well defined. The field-line tracing and interpolation unavoidable adds a certain degree of numerical dissipation -- and since the parallel dynamics are much faster than perpendicular dynamics (flute mode behaviour), this could overwhelm the actual perpendicular dynamics unless special care is taken in the construction of parallel operators (i.e. support operator method). Finally, because the grid is not conformal to the boundary, the application of boundary conditions is complicated significantly.

\begin{figure}[t]
\floatbox[{\capbeside\thisfloatsetup{capbesideposition={right,top},capbesidewidth=0.5\textwidth}}]{figure}[\FBwidth]
{\includegraphics[height=6.1cm,keepaspectratio, trim={0.3cm 0.8cm 0.2cm 0.2cm}]{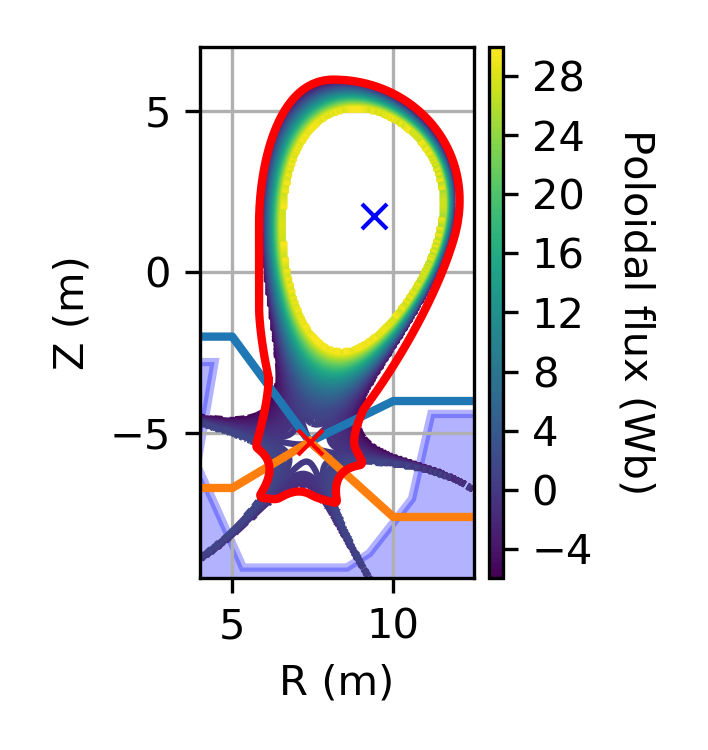}}
{\caption{Preprocessed Snowflake magnetic equilibria. Contours show the given poloidal flux $\Psi$ (in Wb), and the red polygon indicates the first-wall plus divertor. Additionally, the blue \& orange lines gives the upper edge of polygons which provide region-localised flux limiting to eliminate the `dome' parallel boundary, the blue-shaded region gives the end-of-domain boundary, and the blue and red crosses give the magnetic axis and primary x-point respectively. In this figure values of $\Psi$ which are outside of the identified flux-limits are excluded, so the contour region is approximately equal to the resulting numerical grid.}
  \label{fig:SF_preprocessing}}
\end{figure}

\section{Handling of numerical geometries}

\begin{figure}[b]
    \floatbox[{\capbeside\thisfloatsetup{capbesideposition={right,top},capbesidewidth=0.4\textwidth}}]{figure}[\FBwidth]
    {\includegraphics[height=4.5cm,keepaspectratio,trim={4cm 1cm 2cm 0.5cm}]{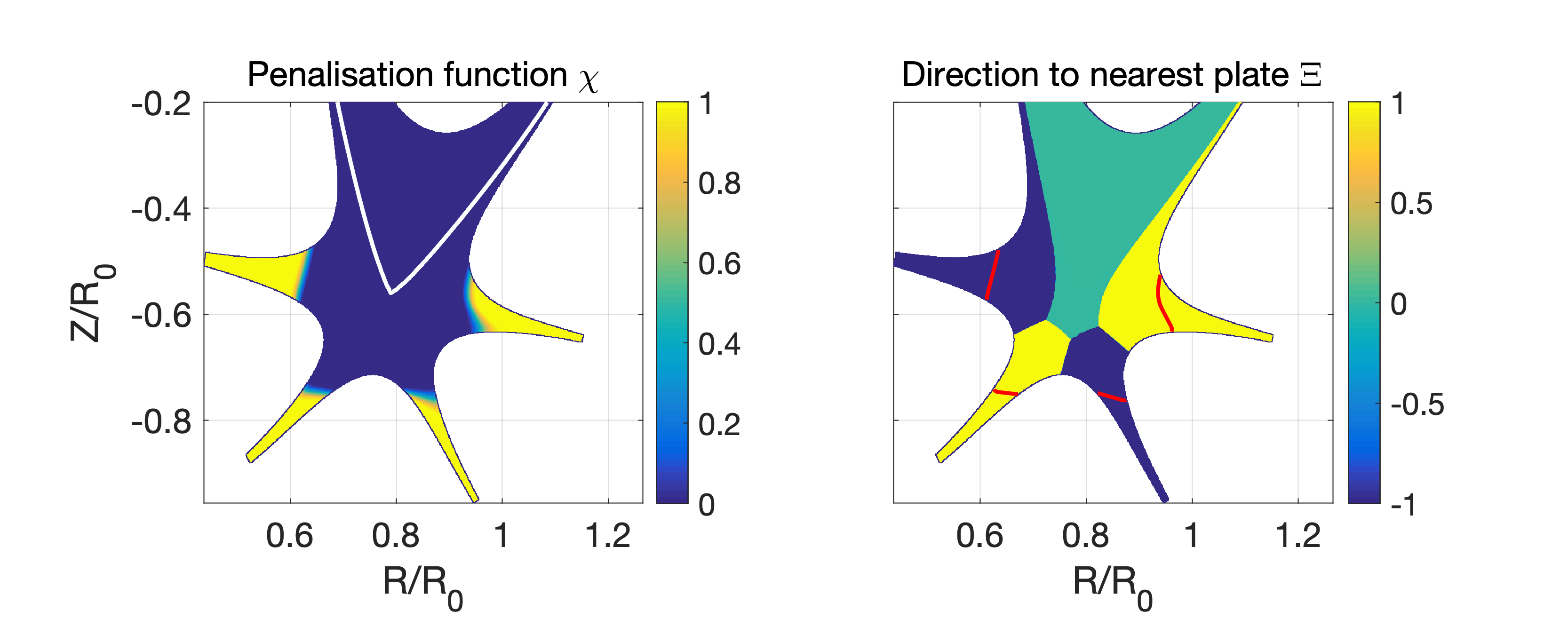}}
    {\caption{Penalisation functions for the Snowflake divertor, showing only the divertor region. The $\chi$ function (\textit{left}) equals $1$ for points inside the $\mathcal{P_\text{div}}$ divertor polygon, and $0$ for points outside, with a smooth transition between these regions. The $\Xi$ (\textit{right}) function gives $+1$ when the nearest plate is in the positive trace direction, and $-1$ if it is in the negative trace direction. For points very far from either plate, the function returns $0$. The plots are annotated with the separatrix (white) for $\chi$ and divertor (red) for $\Xi$.}
    \label{fig:penalisation}}
\end{figure}

\subsection{Preprocessing}
The advanced divertor configurations were supplied as \textit{eqdsk}-standard files, which give the poloidal flux function $\Psi$ as a function of the radial $R$ and vertical $Z$ directions. Additionally, the first-wall and divertor were given as a set of $(R,Z)$ polygon-points. Before importing this data into GRILLIX, a preprocessing step was used to identify features to make the equilibria compatible with the set of equations and boundary conditions in GRILLIX -- for the Snowflake scenario these are indicated in figure \ref{fig:SF_preprocessing}, including \textit{a) a lower limit of $\Psi$}, below which grid points are not included in the computational grid -- excluding the core where the fluid model is not valid; \textit{b) an upper limit of $\Psi$} which eliminates points which are on a flux surface which impacts the first-wall -- giving parallel boundary conditions only at the divertor targets; \textit{c) region-specific $\Psi$ limits \& corresponding region polygons}, which eliminate points which are on a flux-surface which impacts the `dome' region between two divertor targets; \textit{d) a polygon identifying the end of the domain} -- where simple boundary conditions are applied, to provide an end-of-domain for the penalisation stencil; and \textit{e) positions of a primary X-point at $(R_X, Z_X)$ and the magnetic axis at $(R_O, Z_O)$} -- used for normalisation.

\subsection{Poloidal flux $\Psi$ and first-wall/divertor polygon $\mathcal{P_\text{div}}$}
The equilibrium data and results from preprocessing are passed to GRILLIX and used to build a bicubic spline interpolator\cite{bspline-fortran}. This allowed for the poloidal flux-function to be determined at any point within the computational grid, as well as the derivatives of $\Psi$ at those points. Due to the axisymmetry of the magnetic field, the field components are defined entirely by the poloidal flux function via $B_Z(R,Z) = \frac{1}{2\pi R} \frac{\partial \Psi (R,Z)}{\partial R}$ and $B_R(R,Z) = -\frac{1}{2\pi R} \frac{\partial \Psi (R,Z)}{\partial Z}$ which automatically fulfils $\nabla\cdot\mathbf{B} = 0$ \cite{zohm_magnetohydrodynamic_2014}.

The first-wall and divertor polygon, the flux-limiting regions and the end-of-domain polygon are treated via a custom polygon object, which stores the $(R,Z)$ array of polygon points and uses the `winding algorithm' \cite{kumar_extension_2018} to identify whether a query point lies within the interior or exterior of the polygon. Together, the $\Psi$ interpolator and the polygons are used to determine the computation grid and ghost points from a full Cartesian grid.

\subsection{Field-line tracing}
Once the grid has been established, the parallel operators are constructed via field-line tracing. For a given start point $(R_i, Z_i, \phi_i)$ (where $\phi$ is the toroidal angle coordinate), the coordinate variation along the fieldline is given by $\frac{d R}{d \phi} = \frac{B_R}{B_\phi}$ and $\frac{d Z}{d \phi} = \frac{B_Z}{B_\phi}$ and so the fieldline trace from a toroidal angle $\phi_i$ to $\phi_f = \phi_i + \Delta_\phi$ can be expressed as $R_f(\phi_f) = R_i + \int_{0}^{\Delta\phi} \frac{B_R}{B_\phi} d\phi$ and $Z_f(\phi_f) = Z_i + \int_{0}^{\Delta\phi} \frac{B_Z}{B_\phi} d\phi$.

The integration is performed using the \texttt{dop853} integration library\cite{dop853}. The integrator uses an 8th-order Runga-Kutta method with a 5th order error estimator (plus a 3rd order `dense output' interpolation). The error estimator can be used to automatically adapt the step-size of the integrator to achieve a specified tolerance. By adding a modified `trace' toroidal angle $\phi^*$ to the state vector, the adaptive step-size routine (combined with a stop condition) may be used to perform an accurate conditional trace. A \textit{interior trace} of $\phi^*$ may be constructed by checking whether the state vector coordinates are inside or outside of the first-wall/divertor polygon $\mathcal{P_{\text{div}}}$, and `integrating' $\phi^*$ with $\frac{d \phi^*}{d \phi} = \begin{cases} 1 \text{ if ($R',Z' \in \mathcal{P_{\text{div}}}$)} \\ 0 \text{ otherwise} \end{cases}$. Once the point crosses out of the first-wall/divertor polygon, the integration is aborted. A corresponding \textit{exterior trace} can be constructed from the inverse integration condition. This method allows the determination of the trace angle $\phi^*$ to the divertor targets in the direction with or against the toroidal field, marked as $\phi_{+,int} > 0$ and $\phi_{-, int} <0$ respectively. For points outside of the vessel, only one of these angles will be defined. The external trace angles are defined as $\phi_{\pm,ext}=\begin{cases}\phi_{\pm,ext}\pm\phi_\Sigma \text{ if } \exists \phi_{\pm,ext} \\ \phi_{\mp,ext} \text{ if } \exists \phi_{\mp,ext}\end{cases}$ (for $\phi_\Sigma > 0$ the interior trace angle between the divertor targets) which gives continuous functions $\phi_+$ and $\phi_-$. At the targets, $\phi_\pm = \pm\phi_\Sigma$ and $\phi_\mp = 0$ if the field line is directed towards the target, and vice versa.

\begin{figure}
    \floatbox[{\capbeside\thisfloatsetup{capbesideposition={right,top},capbesidewidth=0.45\textwidth}}]{figure}[\FBwidth]
    {\includegraphics[height=8cm,keepaspectratio, trim={3cm, 0.8cm, 3cm, 1cm},clip]{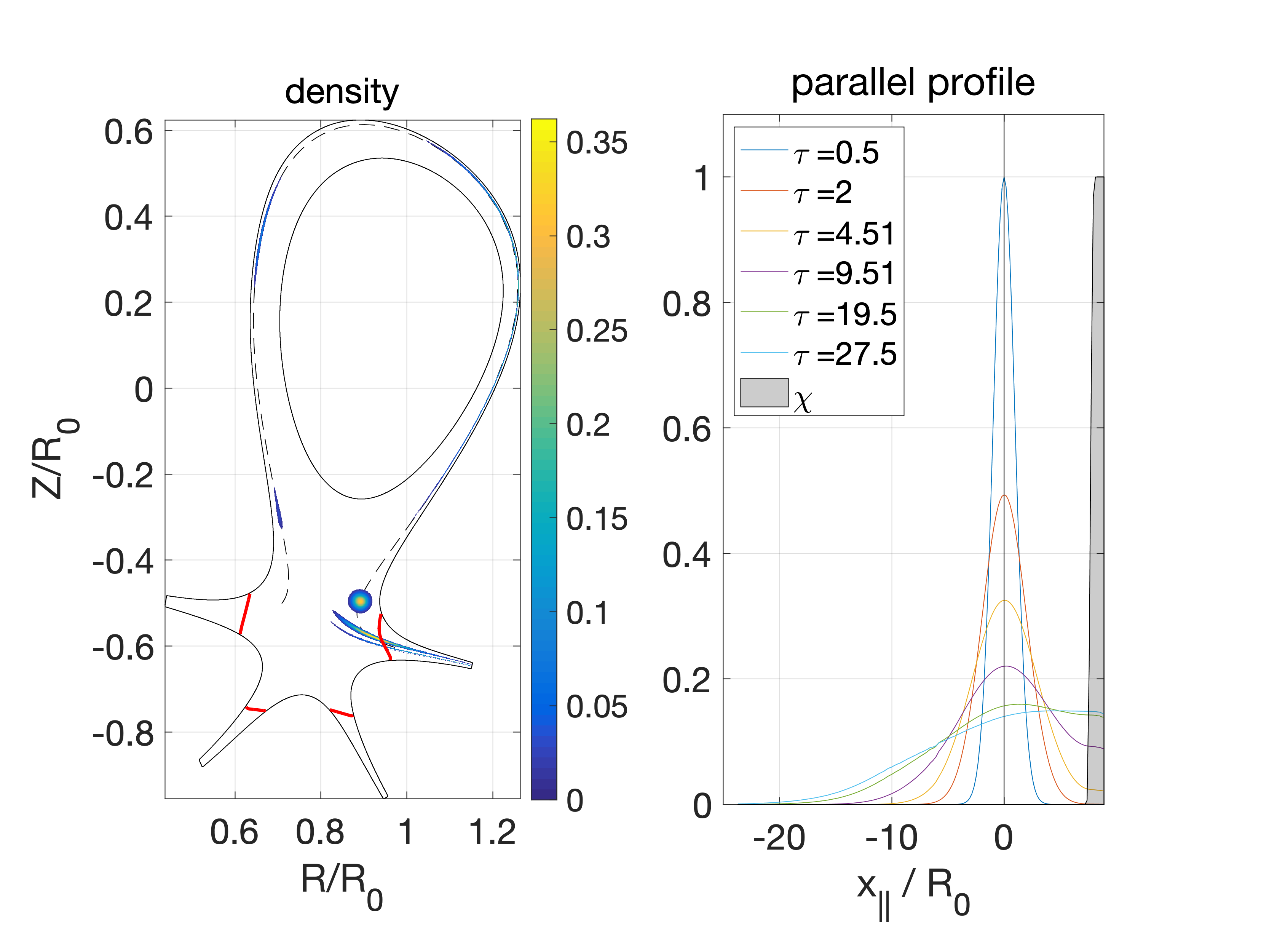}}
    {\caption{Parallel diffusion case, for a blob initiated at $R/R_0 = 0.89, Z/R_0 = -0.495$ on a single plane with an initial density of $1$. The \textit{left} image shows the density distribution on a single poloidal plane after $\tau = 27.5$, as well as the magnetic field-line corresponding to the centre of the blob (black-dashed line). The initial circular blob can be seen in its original position, as well as strongly sheared blobs corresponding to $(-3\leq n \leq 2)\times2\pi$ full-rotation projections of the blob. The $(n<-3)$ and $(n>3)$ projections are in the penalisation area, and so are not visible. The \textit{right} plot shows the density profile of the blob in the parallel direction (with $n(\tau=0.5)$ set equal to $1$), as well as the penalisation $\chi$ function (black dashed line). It is seen that the initial density evolution is a symmetric decay. Once the forward density diffusion front reaches the boundary condition (where $\chi = 1$), the Neumann boundary condition modifies the dynamics.}\label{fig:SF_diffusion}}
\end{figure}

\subsection{Penalisation functions}
Parallel boundary conditions in GRILLIX are applied via the \textit{penalisation method} (see section III.B. of Stegmeir, A. et al., 2019 \cite{stegmeir_global_2019}), which allows for the treatment of arbitrarily complex boundary structures (limited only by the grid resolution). For some variable $f(R,Z,\phi,t)$ with a dynamical forcing term $F(R,Z,\phi,t)$, the interior evolution of the variable within the domain $\Omega$ is given by the differential equation $\frac{\partial f(R,Z,\phi,t)}{\partial t} = F(R,Z,\phi,t)$ for $(R,Z) \in \Omega$. Rather than directly forcing the parallel ghost cells to fulfil the boundary conditions specified, a modified equation is solved over union of the interior domain $\Omega$ and the parallel ghost cells $\Omega_\parallel$, such that
$$\frac{\partial f(R,Z,\phi,t)}{\partial t} = (1-\chi(R,Z))F_f(R,Z,\phi,t) + \frac{\chi(R,Z)}{\epsilon}(f_P(R,Z,\phi,t) - f(R,Z,\phi,t)) \text{ for } (R,Z) \in \Omega \cup \Omega_\parallel$$
Here $\chi(R,Z) = \begin{cases}0 \text{ if } (R,Z) \in \Omega\\ 1 \text{ if } (R,Z) \in \Omega_\parallel\end{cases}$ is the characteristic function, $\epsilon \ll 1$ is the relaxation parameter, and $f_P$ is a penalisation value which is set such that the boundary conditions are satisfied exactly as $\epsilon \to 0$. Within GRILLIX, the toroidally-projected width of the $\chi$ function must be greater than or equal to the toroidal angle between poloidal planes. This prevents an issue where the staggered and full-grids used for the finite-volume discretisation of parallel dynamics decouple at the boundary. By using a finite width and gradually transitioning from interior to exterior (penalised) dynamics, this decoupling can be avoided.

The smoothed penalisation characteristic function can be determined directly from the trace angles $\phi_+$ and $\phi_-$ via $\chi = 1 + \mathcal{S}_3(\phi_-, \chi_W) - \mathcal{S}_3(\phi_+, \chi_W)$. Here $\mathcal{S}_3$ is a 3rd order \textit{smooth-step} function -- a sigmoid function which gives a smooth (Hermite-polynomial) transition from 0 to 1, over a region of width $\chi_W$. Additionally, a function $\Xi = 1 - 2\cdot\mathcal{H}(\phi_+ + \phi_-)$ (where $\mathcal{H}$ is a Hermite step-function) is developed to indicate, for a given point, whether the nearest plate is in the direction of ($\Xi = +1$), or against ($\Xi = -1$), the toroidal field direction. This is required to indicate the sign for the penalisation values of the parallel velocity and parallel current, as well as to determine which direction is `upstream' (towards the interior) for boundary conditions which set ghost values in terms of upstream neighbours (i.e. $\geq 1^\text{st}$ order Dirichlet, Neumann and higher derivatives). These functions are shown for the Snowflake divertor in figure \ref{fig:penalisation}.

The use of a smooth $\chi$ function is required for numerical stability in GRILLIX. However, this has the disadvantage that the boundary condition is not applied exactly at the plate, but rather the system of equations smoothly transitions to the boundary condition. This complicates the determination of boundary fluxes, and -- in cases where the forcing terms of each variable are highly disparate -- may result in the `boundary' being located at different points for different quantities. As such, choosing a penalisation width as small as possible while still avoiding poloidal plane decoupling is desirable. Future work aims to consider whether other parallel stencils such as WENO could alleviate this issue, or alternatively whether an alternative set of non-decoupling boundary conditions \cite{loizu_boundary_2012} could be made compatible with the implemented turbulence model. Of particular interest is the set of `strong-sink boundary conditions' developed by Parades et al. \cite{paredes_penalization_2014}, which could allow for a unified set of boundary conditions for both the divertor and first wall.

\begin{figure}
    {\includegraphics[height=8cm,keepaspectratio,trim={0cm, 0.1cm, 0cm, 1cm},clip]{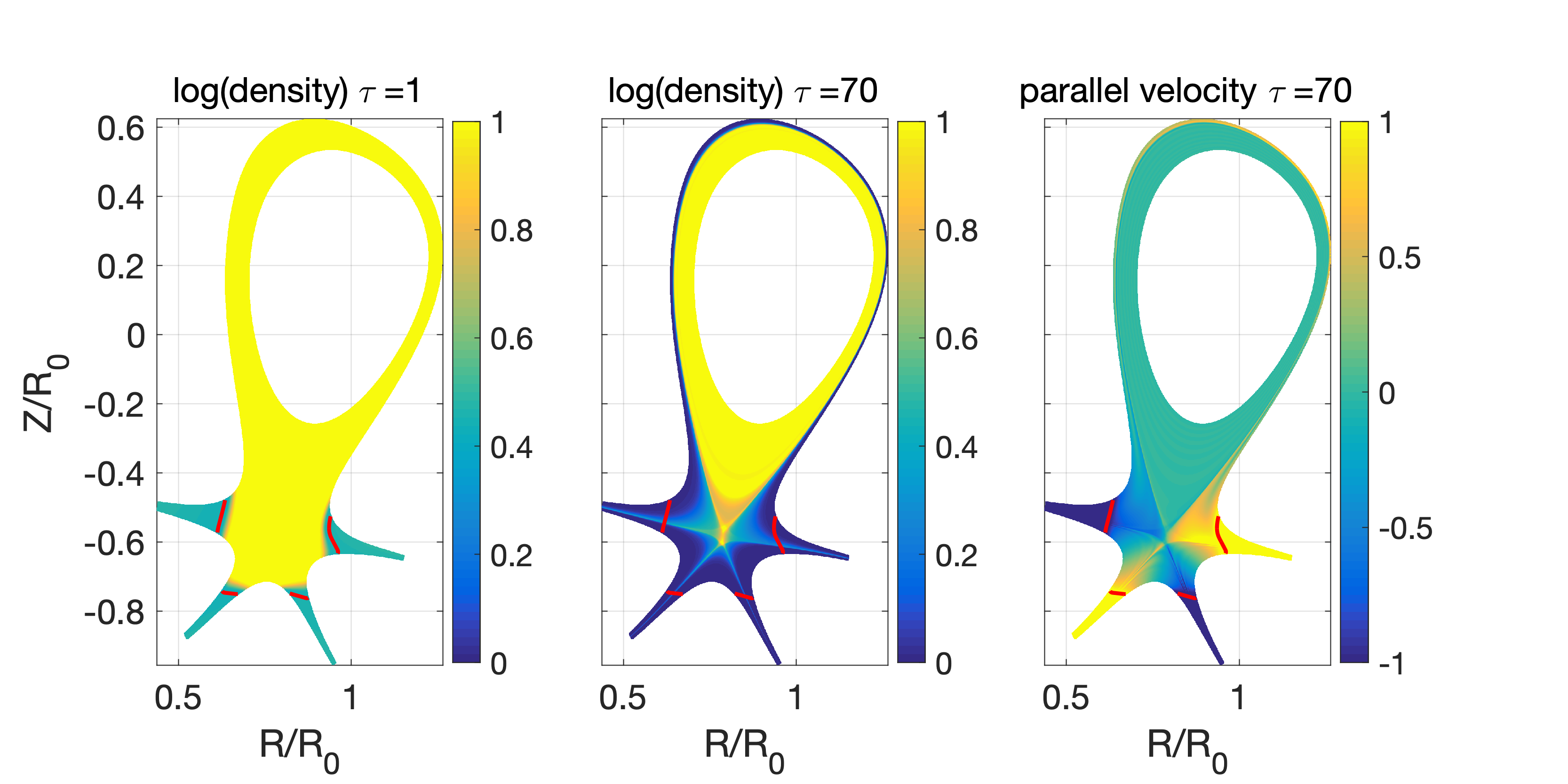}}
    {\caption{Logarithmic-density \textit{(left)} and parallel velocity \textit{(right)} of a Snowflake scenario initialised with $log(n)=1$ and $u_\parallel = 0$ initial conditions, and $u_\parallel = \pm 1$ and $\nabla_\parallel log(n) =0$ boundary conditions, after $\tau = 70$ of evolution. The divertor plates are highlighted in red. It is seen that the density is advected out of the SOL, while within the closed-fieldline region it is relatively static, with density loss near the X-point possibly indicating an ergodic region. The formation of a primary and secondary density accumulation is observed, respectively corresponding to the X-point and poloidal flux plateau visible in fig \ref{fig:SF_preprocessing}. A complex parallel velocity pattern is developed in region near the X-point, with primary flow-channels aligned to the separatrix and the SOL/private-flux bounding field-line.}
    \label{fig:SF_advection}}
\end{figure}

\section{Testing of numerical equilibria}
The physical model in GRILLIX is expressed as a coupled set of non-linear partial differential equations (see \cite{stegmeir_global_2019}), which may be broadly defined as either \textit{parabolic}, \textit{hyperbolic} or \textit{elliptic} PDEs, or a mix thereof. We consider prototypical parabolic and hyperbolic equations -- the parallel diffusion and parallel advection equations -- to test the implementation of the numerical equilibria in GRILLIX.

The parallel advection equation is given by $\partial_t n + \nabla \cdot (n u_\parallel \hat{b}) = 0 \text{ and } \partial_t u_\parallel + u_\parallel \nabla_\parallel u_\parallel = -\frac{\nabla_\parallel n}{n}$. The density $n$ is located on the full-grid $G$, while the parallel velocity is located on the staggered grid $G^*$. These grids are coupled via the parallel operators $\nabla_\parallel(...): G \to G^*$ which maps from full to staggered grid, and $\nabla\cdot(\hat{b}...): G^* \to G$ which maps from staggered to full. The boundary conditions applied are \textit{Bohm} for the parallel velocity ($u_\parallel \to \pm c_s = \pm 1$), and Neumann for the density ($\nabla n = 0$). The time normalisation $\tau$ is equal to $\frac{R_0}{c_s}$ for this system of equations. The results of the advection test are shown in figure \ref{fig:SF_advection}, and discussed in the corresponding caption.

The parallel diffusion equation is $\partial_t n = C \nabla\cdot(\hat{b}\nabla_\parallel n)$, with Neumann density boundary conditions. The time normalisation $\tau$ for this equation is $\frac{R_0^2}{C}$. The results from the diffusion test are shown in figure \ref{fig:SF_diffusion}, and discussed in the corresponding caption.

\section{Conclusions and future work}
Advanced divertor configurations aim to reduce the heat flux impacting the divertor targets via increased flux expansion \& parallel connection lengths, and improved control of detachment. To assist with the interpretation of experimental results, modelling with turbulence codes is desirable. The flux-coordinate independent approach is expected to be well-suited to studying complex magnetic geometries due to the ability to set the grid independent of the magnetic field structure. However, there are also challenges associated with this method -- particularly the application of boundary conditions for boundaries which are non-conformal to the grid. In this paper, we discuss an extension of the penalisation method discussed in \cite{stegmeir_global_2019}, which allows for minimisation of the penalisation-width required for coupling the staggered and full-grids at the boundary.

Additionally, an automatic method for determining the direction to the nearest plate based on a conditional field-line tracing is introduced. The modifications to the code are applied to prototypical parabolic (parallel-advection) and hyperbolic (parallel-diffusion) test cases, which represent the two classes of differential equation which are affected by the magnetic field structure through the parallel operators. The parallel diffusion case is seen to decay as expected initially and is modified by the Neumann boundary condition for late-stage evolution. Blob projections are visible at $n(2\pi)$ projections along the field-line. The separatrix and X-points formation is clearly seen in the advection test case.

Work to apply the full turbulence model on an ADC case is ongoing, with no saturated results available yet.

\subsection*{Acknowledgements}
The authors are grateful to Dr N. R. Walkden for sharing the \texttt{PyEquilibrium} code with them, which greatly helped with the processing of \texttt{geqdsk} files.\\
  
\noindent This work has been carried out within the framework of the EUROfusion Consortium and has received funding from the Euratom research and training programme 2014-2018 and 2019-2020 under grant agreement No 633053. The views and opinions expressed herein do not necessarily reflect those of the European Commission.

\bibliographystyle{wileyNJD-ACS}
{\footnotesize
\setlength{\bibsep}{1pt plus 0.3ex}
\bibliography{PET.bib}
}

\end{document}